\documentclass[aps,twocolumn,superscriptaddress,showpacs,preprintnumbers,prb]{revtex4}
\usepackage{graphicx}
\usepackage{amsmath}
\usepackage{nccmath}
\usepackage{txfonts}
\usepackage{bm} 

\newcommand{\msr}{$\mu$SR}
\newcommand{\lvo}{LiV$_2$O$_4$}

\newcommand{\ymn}{Y(Sc)Mn$_2$}
\newcommand{\ysmn}{Y$_{1-x}$Sc$_x$Mn$_2$}
\newcommand{\ymnz}{YMn$_2$Zn$_{20-x}$In$_x$}
\newcommand{\ymzd}{YMn$_{2+\delta}$Zn$_{20-\delta-x}$In$_x$}
\begin{document}

\preprint{arXiv}

\title{Quest for the Origin of Heavy Fermion Behavior in $d$-Electron Systems}
\author{M. Miyazaki}\thanks{Present address: Muroran Institute of Technology, Muroran, Hokkaido 050-8585, Japan}
\affiliation{Condensed Matter Research Center, Institute of Materials Structure Science, High Energy Accelerator Research Organization (KEK), Tsukuba, Ibaraki 305-0801, Japan}
\affiliation{Muon Science Laboratory, Institute of Materials Structure Science, High Energy Accelerator Research Organization (KEK), Tsukuba, Ibaraki 305-0801, Japan}
\affiliation{Department of Materials Structure Science, The Graduate University for Advanced Studies, Tsukuba, Ibaraki 305-0801, Japan}
\author{I. Yamauchi}\thanks{Present address:  Faculty of Science and Engineering, Saga University, Saga 840-8502, Japan}
\affiliation{Condensed Matter Research Center, Institute of Materials Structure Science, High Energy Accelerator Research Organization (KEK), Tsukuba, Ibaraki 305-0801, Japan}
\affiliation{Muon Science Laboratory, Institute of Materials Structure Science, High Energy Accelerator Research Organization (KEK), Tsukuba, Ibaraki 305-0801, Japan}
\affiliation{Department of Materials Structure Science, The Graduate University for Advanced Studies, Tsukuba, Ibaraki 305-0801, Japan}
\author{R. Kadono}\thanks{Corresponding author: ryosuke.kadono@kek.jp}
\affiliation{Condensed Matter Research Center, Institute of Materials Structure Science, High Energy Accelerator Research Organization (KEK), Tsukuba, Ibaraki 305-0801, Japan}
\affiliation{Muon Science Laboratory, Institute of Materials Structure Science, High Energy Accelerator Research Organization (KEK), Tsukuba, Ibaraki 305-0801, Japan}
\affiliation{Department of Materials Structure Science, The Graduate University for Advanced Studies, Tsukuba, Ibaraki 305-0801, Japan}

\begin{abstract}
Spin fluctuation is presumed to be one of the key properties in understanding the microscopic origin of heavy-fermion-like behavior in the class of transition-metal compounds, including \lvo, \ymn, and YMn$_2$Zn$_{20}$. In this review, we demonstrate by our recent study of muon spin rotation/relaxation that the temperature ($T$) dependence of the longitudinal spin relaxation rate ($\lambda\equiv1/T_1$) in these compounds exhibits a common trend of leveling off to a constant value ($\lambda\sim const$.) below a characteristic temperature, $T^*$.  This is in marked contrast to the behavior predicted for normal metals from the Korringa relation, $\lambda\propto T/\nu$, where the spin fluctuation rate ($\nu$) in the Pauli paramagnetic state is given as a constant, $\nu\simeq 1/[h D(E_F)]$ [with $D(E_F)$ being the density of states at the Fermi energy]. Thus, the observed behavior of $\lambda$ implies that the spin fluctuation rate becomes linearly dependent on temperature, $\nu\propto T$, suggesting that heavy quasiparticles develop in a manner satisfying $D(E_F)\propto (m^*)^{\sigma}\propto 1/T$ at lower temperatures ($\sigma$ determined by the electronic dispersion). Considering that the theory of spin correlation for intersecting Hubbard chains as a model of pyrochlore lattice predicts $\nu\propto T$, our finding strongly indicates the crucial role of $t_{2g}$ bands which preserve the one-dimensional character at low energies due to the geometrical frustration specific to the undistorted pyrochlore lattice.
\end{abstract}
\pacs{71.27.+a, 75.20.Hr, 76.75.+i}

\maketitle

\section{Introduction}

Geometrical frustration in electronic degrees of freedom such as spin, charge, and orbit, which is often realized in the stages of highly symmetric crystals, has been one of the major topics in the field of condensed matter physics. In particular, the heavy fermion (HF) behavior in \ysmn\ [\ymn]\cite{Wada:89,Fisher:93} and \lvo\ \cite{Kondo:97,Urano:00} has attracted broad interest, where such a local electronic correlation specific to the highly symmetric pyrochlore structure may be of direct relevance to the formation of the heavy quasiparticle (QP) state. However, despite decades of studies, the microscopic mechanism by which the local correlation is transformed into the heavy QP mass of itinerant $d$-electrons in these compounds still remains controversial.

In general, the development of heavy QPs accompanies narrowing of the effective band width ($W$) or an increase in the density of states (DOS) at the Fermi energy [$D(E_F)$].
It is thus expected from a naive consideration based on Heisenberg's uncertainty principle that the HF behavior should manifest itself in the spin dynamics as reduction of the spin fluctuation rate ($\nu$) because these quantities are mutually linked by the following relation:
\begin{equation}
\nu\sim \frac{W}{h}\simeq\frac{1}{hD(E_F)}.
\end{equation}
Since the effective QP mass is directly connected to the DOS via the relation $m^*\propto [D(E_F)]^{2/3}$ for a three-dimensional (3D) Fermi gas, studies on the spin fluctuation with particular emphasis on relatively low energies should provide valuable information on the mechanism of heavy QP formation.

As a probe of spin fluctuation, muon spin rotation (\msr) has a unique frequency window with high sensitivity for $10^5\le\nu\le 10^{11}$ s$^{-1}$,  filling the gap between those covered by nuclear magnetic resonance (NMR) and neutron scattering.  This provides promising perspectives for the $\mu$SR study of $d$-electron HF-like compounds. Here, we establish that the longitudinal spin relaxation rate ($\lambda\equiv1/T_1$)  in the above-mentioned compounds exhibits a common feature that it asymptotically becomes independent of temperature upon cooling below a crossover temperature of $T^*\simeq10^1$--$10^2$ K.  This, within the framework of fermionic QPs, means that the spin fluctuation rate becomes linearly dependent on temperature, $\nu\propto T$, which is in marked contrast to ordinary metals, where  $\nu$ is independent of $T$ as determined by the DOS at the Fermi level. 
Considering that the transition-metal ions  in these compounds comprise a pyrochlore lattice, it is naturally expected that the above-mentioned feature in the spin dynamics will be specific to the relevant lattice structure.  We argue that this is indeed the case in that the underlying spin dynamics can be understood by the spin correlation of the intersecting Hubbard chains, which provides a model of the pyrochlore lattice. This strongly indicates the crucial role of $t_{2g}$ bands that preserve the one-dimensional (1D) character specific to the undistorted pyrochlore lattice at low energies, which is consistent with a theoretical scenario that the 1D-3D crossover due to coupling between these 1D chains developing at low temperatures is the origin of the HF behavior.

In the following, we present an overview of our \msr\ results for \ymn\cite{Miyazaki:11}, \ymnz,\cite{Miyazaki:14} and \lvo,\cite{Koda:04,Kadono:12} where the temperature dependence of $\lambda$ is the primary focus.  Then, the temperature dependence of the spin fluctuation rate is discussed in connection with a possible scenario for the origin of heavy QPs in these compounds.
We also show that the anomalous behavior of $\lambda$ is accompanied by strong broadening of the linewidth under a transverse field ($\lambda_\perp\equiv1/T_2$), which is not explained by the hyperfine parameters extrapolated from their values at high temperatures\cite{Koda:05,Yamauchi:14}.

\section{Muon Spin Relaxation in Itinerant Electron Systems}
As a probe of spin fluctuation in condensed matter, \msr\ is kin to NMR, primarily because of the fact that an implanted muon can be regarded as a light radioisotope of a proton ($^1$H).
Nonetheless, while the physics behind the mechanism of spin relaxation is common, there are two major factors, i.e., the time windows of observation (10$^{-4}$--10$^{0}$ s for NMR, 10$^{-9}$--10$^{-5}$ s for \msr) and the relevant sites (on-site for NMR, interstitial site for \msr), that make these two techniques distinct (or even complementary with each other).   It would be useful to highlight such differences by examining the longitudinal spin relaxation rate ($1/T_1$) in simple metals.  

According to Korringa's relation, the spin relaxation induced by $s$ band electrons is given by 
\begin{equation}
T_1TK_s^2=\frac{\hbar}{4\pi k_B}\left(\frac{\gamma_e}{\gamma_\mu}\right)^2,\label{tnt}
\end{equation}
where $\gamma_e$  ($=2\pi\times28024$ MHz) and $\gamma_\mu$ ($=2\pi\times135.53$ MHz) are the respective gyromagnetic ratios of an electron and muon. 
$K_s$ is the Knight shift, which is expressed in terms of the hyperfine field ($H_s$) per electron as
\begin{equation}
K_s=\frac{H_s}{N_A\mu_B}\chi_{\rm p},
\end{equation}
$N_A$ is Avogadro's number,  $\mu_B$ is the Bohr magneton, and $\chi_{\rm p}$ is the magnetic susceptibility [$\chi_{\rm p}=2\mu_B^2D(E_F)$ for uncorrelated electrons].  Equation (\ref{tnt}) may be rewritten as
\begin{equation}
\lambda\equiv\frac{1}{T_1}=\frac{4\pi}{\hbar}(\hbar\gamma_\mu H_\alpha)^2\left[\frac{D_\alpha(E_F)}{N_A}\right]^2k_BT,
\end{equation}
where $H_\alpha$ is the hyperfine field of electrons at $\alpha$ orbitals ($\alpha=s,\:p,\:d,...$) with $D_\alpha(E_F)$ denoting the DOS for the corresponding band.

Here, let us consider the example of pure silver (Ag), in which the muon Knight shift has been reported to be $K_s=94(3.5)$~ppm~\cite{Schenck}.  Equation (\ref{tnt}) then yields
$$T_1T\equiv\frac{T}{\lambda}\simeq 2.94\:\: [{\rm K}\cdot{\rm s}].$$
 A similar estimation for silver nuclei (e.g., $^{107}$Ag, where $K_s\simeq0.52$\%) leads to $T_1T\simeq 5.93$ [K$\cdot$s], which is comparable to the $T_1T$ value for muon.
Considering the melting point of silver ($\simeq10^3$ K), it implies that $1/T_1\le10^3$ s$^{-1}$, which is far below the sensitive range for \msr, while it is readily observed by NMR.  Thus, the $s$ band electron makes a negligible contribution to $1/T_1$ in \msr. 

When the electronic correlation is not negligible, it is useful to resort to a more general form of $1/T_1$ with the relevant spin fluctuation expressed by spin density operators,
\begin{equation}
\lambda=\gamma_\mu^2\sum_{\bm q}A_{\bm q}A_{-{\bm q}}\int_{-\infty}^{\infty}dt\cos\omega_\mu t\frac{\langle [S^+_{\bm q}(t),S^-_{-{\bm q}}(0)]\rangle}{2},\label{lmdsq}
\end{equation}
where $\omega_\mu=\gamma_\mu B_0$ ($B_0$ being the external field) and $A_{\bm q}$ and $S^\pm_{\bm q}$ are the Fourier components of the hyperfine field [$A_\mu({\bm r})$] and spin density [$S^+_{\bm q}=\sum_{\bm k}c^*_{{\bm k}+{\bm q},\uparrow}c_{{\bm k},\downarrow}$, $S^-_{\bm q}=\sum_{\bm k}c^*_{{\bm k}+{\bm q},\downarrow}c_{{\bm k},\uparrow}$, with $c^*_{{\bm k},\uparrow}/c_{{\bm k},\downarrow}$ being the creation/annihilation operators for spin up($\uparrow$)/down($\downarrow$) electrons], respectively, $[A,B]=\frac{1}{2}(AB+BA)$, and $\langle Q\rangle={\rm Tr}[e^{-(H'/k_BT)Q}]/{\rm Tr}[e^{-(H'/k_BT)}]$ (with $H'$ being the Hamiltonian of the electron system) \cite{Moriya:56}. 
Imposing the fluctuation-dissipation theorem, which links the thermal fluctuation described by the correlation function to the imaginary part (dissipation) of the generalized susceptibility,
\begin{equation}
\frac{2\hbar\chi^{''}_\perp({\bm q},\omega_\mu)}{N_A\mu_B^2(1-e^{-\hbar\omega_\mu/k_BT})}
=\int_{-\infty}^\infty dt\cos\omega_\mu t\frac{\langle [S^+_{\bm q}(t),S^-_{-{\bm q}}(0)]\rangle}{2},
\end{equation}
to Eq.~(\ref{lmdsq}) with the further assumption that $\hbar\omega_\mu\ll k_BT$ leads to
\begin{equation}
\lambda\simeq \frac{2\gamma_\mu^2}{N_A\mu_B^2}k_BT\sum_{\bm q}A_{\bm q}A_{-{\bm q}}\frac{\chi^{''}_\perp({\bm q},\omega_\mu)}{\omega_\mu},\label{lmdq}
\end{equation}
where $\chi^{''}_\perp({\bm q},\omega_\mu)$ is the imaginary part of the dynamical spin susceptibility (perpendicular to the quantization axis that is usually determined by $B_0$).

 The actual form of $A_{\bm q}$ is determined by $A_\mu({\bm r})$  (with ${\bm r}=0$ at the muon position), and thus it generally depends on the muon site(s). An exception is the case that the hyperfine field is predominantly determined by the Fermi contact interaction, $A_\mu({\bm r})\simeq A_\mu\cdot\delta({\bm r})$, so that $A_{\bm q}=A_\mu$ irrespective of ${\bm q}$. Equation (\ref{lmdq}) is then reduced to
\begin{equation}
\lambda\simeq \frac{2\gamma_\mu^2}{N_A\mu_B^2}k_BT A_\mu^2\sum_{\bm q}\frac{\chi^{''}_\perp({\bm q},\omega_\mu)}{\omega_\mu}.\label{lmdchiqz}
\end{equation}

 However, such a situation is rare for \msr\ because of the small charge of a muon (same as that of a proton) and  the fact that muons are usually located at the interstitial sites.  Consequently, the magnetic dipolar interaction is the primary source of hyperfine fields acting on muons, 
 \begin{equation}
 \hat{A}_\mu({\bm r})=A^{\alpha\beta}({\bm r})=\sum_j\frac{1}{r_j^3}\left(\frac{3\alpha_j\beta_j}{r_j^2}-\delta_{\alpha\beta}\right)\:\:(\alpha,\beta=x,y,z),
 \end{equation}
 where ${\bm r}_j=(x_j,y_j,z_j)={\bm R}_j-{\bm r}$ with ${\bm R}_j$ being the position of the $j$th electron.  
More specifically, the effective hyperfine field in the paramagnetic state is given by the second moment 
\begin{equation}
\hat{A}_\mu^2({\bm r})=\sum_{\alpha,\beta} [A^{\alpha\beta}({\bm r})]^2,
\end{equation}
where the corresponding Fourier transform $(A^{\alpha\beta}_{\bm q})^2$ can be numerically evaluated for the specific muon site(s).  An example calculated for ${\bm q}\parallel [110]$ in \ymn\ is shown in Fig.~\ref{ymn-cr}(b), where the muon is presumed to be at the 16$c$ site [see Fig.~\ref{ymn-cr}(a)].  It shows a broad distribution with a peak near ${\bm q}\simeq0$ and a tail over the region $|{\bm q}|\leq$ $\sim$1  \AA$^{-1}$. A similar result is observed for the [1-10] direction, indicating that Eq.~(\ref{lmdq}) can be approximated by the form  
\begin{equation}
\lambda\simeq \frac{2\gamma_\mu^2}{N_A\mu_B^2}k_BT\sum_{\alpha=x,y;\:\beta=x,y,z}(A^{\alpha\beta}_{{\bm q}\simeq0})^2\frac{\chi^{''}_\perp({\bm q}\simeq0,\omega_\mu)}{\omega_\mu}.\label{rfd0}
\end{equation}

Following the assumption generally adopted in NMR that the density spectrum of spin fluctuation is described by the Lorentzian distribution function $f(\omega)$ with its amplitude represented by the local susceptibility $\chi$,
\begin{equation}
\frac{\chi^{''}_\perp({\bm q}\simeq0,\omega_\mu)}{\omega_\mu}\simeq\chi \cdot f(\omega_\mu)=\chi\frac{\nu}{\nu^2+\omega^2_\mu},\label{lrntz}
\end{equation} 
Eq.~(\ref{rfd0}) is further simplified to 
\begin{equation}
\lambda(\nu,\omega_\mu)\simeq  \frac{k_BT\chi}{N_A\mu_B^2}\cdot\frac{(\delta_\mu^\parallel)^2\nu}{\nu^2+\omega_\mu^2},\label{rfd}
\end{equation}
where $\delta_\mu^\parallel$ is the redefined hyperfine parameter and $\nu$ is the fluctuation rate of $\delta_\mu^\parallel$  at ${\bm q}\simeq0$. The above equation is valid when $\nu\gg\delta_\mu^\parallel$.  Equation (\ref{lrntz}) corresponds to the general assumption in the time domain that the correlation of a fluctuating hyperfine field can be described by the stationary Gaussian--Markovian process in the approximated form
\begin{equation}
\langle A_\mu(t_0)A_\mu(t_0+t)\rangle=\langle A_\mu^2\rangle\exp(-\nu|t|).
\end{equation} 

While \msr\ and NMR can be used to observe the spin fluctuation via the same process described by Eq.~(\ref{rfd}), their sensitive ranges are markedly different (as already discussed in the case of $s$ band electrons).  As illustrated in Fig.~\ref{ymn-cr}(c), the difference stems from the sensitive range of $\lambda$ ($=1/T_1$), which is determined by the accessible time window, $T_{\rm W}$, for the respective probes (see also Table \ref{tab}). The present study benefits greatly from this unique sensitive range of \msr\ to spin fluctuation.  
We rely on Eq.~(\ref{rfd}) to deduce $\nu$ explicitly from the experimentally determined longitudinal spin relaxation rate $\lambda$. 
\begin{table}[ht]
\begin{tabular}{c|cc}
\hline\hline
 & \msr\ & NMR \\
 \hline
 Time window ($T_{\rm W}$) & $10^{-9}\le T_{\rm W}\le10^{-5}$ [s] & $10^{-4}\le T_{\rm W}\le10^{1}$ [s]\\
Fluctuation rate ($\nu$) & $10^4\le\nu\le10^{11}$ [s$^{-1}$] & $\nu\le10^4$, $\nu\ge10^{11}$ [s$^{-1}$]\\
\hline\hline
\end{tabular}
\caption{Sensitive ranges of spin fluctuation rate for \msr\ and NMR predicted from Eq.~(\ref{rfd}).  See Fig.~\ref{ymn-cr}(c) for more details. }\label{tab}
\end{table}

\section{Overview of $\mu$SR Results}
\subsection{Y(Sc)Mn$_2$}
\begin{figure}[t]
\begin{center}
\includegraphics[width=0.475\textwidth]{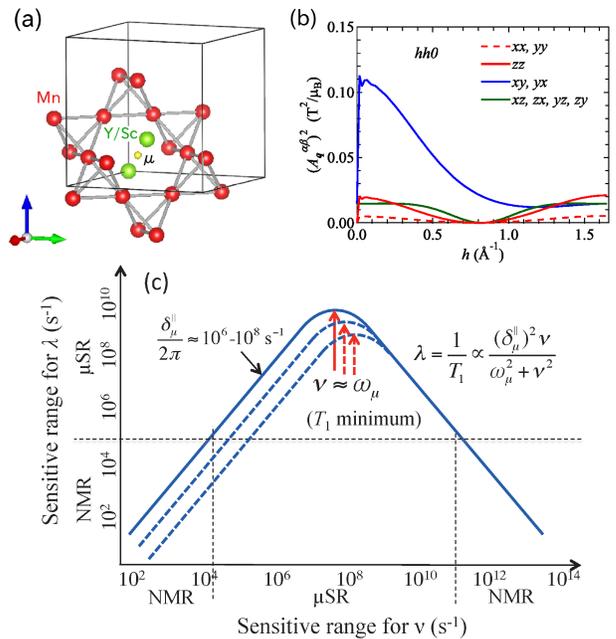}
\caption{(Color online) (a) Crystal structure of \ymn, where Y/Sc and Mn atoms are indicated by green and red spheres, respectively.  Mn atoms form a network of corner-shared tetrahedra known as a pyrochlore lattice. A small yellow circle shows the 16$c$ site presumed to be occupied by implanted muons. (b) Fourier transform of the hyperfine field distribution $\hat{A}^2_\mu({\bm r})$ in \ymn\ with a 16$c$ site at the origin. (c) Sensitive range of spin fluctuation rate for \msr\ and NMR predicted from Eq.~(\ref{rfd}), where the hyperfine field is presumed to be $10^{-2}$--$10^0$ T. Note that the sensitive ranges for NMR are split into two different ranges, where the high-frequency range corresponds to the region close to the limit of motional narrowing. Dashed curves show $\lambda$ for greater external magnetic fields ($\propto\omega_\mu$).}\label{ymn-cr}
\end{center}
\end{figure}

Yittrium manganite (YMn$_2$) is an intermetallic Laves phase (C15-type) compound and was the first transition-metal system in which HF behavior was observed. As shown in Fig.~\ref{ymn-cr}, it comprises a 3D network of corner-shared tetrahedra with Mn ions at their corners, resulting in a scheme equivalent to a cubic pyrochlore lattice. Although YMn$_2$ exhibits magnetic order with complex helical modulation and a large volume expansion below $T_{\rm N} \simeq100$ K\cite{Ballou:87}, it remains in the paramagnetic state under hydrostatic pressure ($\ge0.4$ GPa) or upon the substitution of Y by Sc (\ysmn, with $ x\ge 0.03$), which is also accompanied by a large increase in the QP mass ($m^* \simeq15$ times the band mass) as inferred from the electronic specific heat \cite{Fisher:93}. 

Our \msr\ measurements were performed using polycrystalline samples of \ysmn\ with various Sc contents ($x=0.03$, 0.05, 0.07, and 0.08, as prepared) grown from melts in an argon arc furnace followed by annealing, where the details of sample preparation are described elsewhere \cite{Nakamura:88}.  
The bulk properties of these samples including the electronic specific heat (Sommerfeld) coefficient and uniform spin susceptibility ($\chi$) were in good agreement with earlier reports (see Ref.~\cite{Miyazaki:11} for more details). 

\begin{figure}[tb]
\begin{center}
\includegraphics[width=0.375\textwidth]{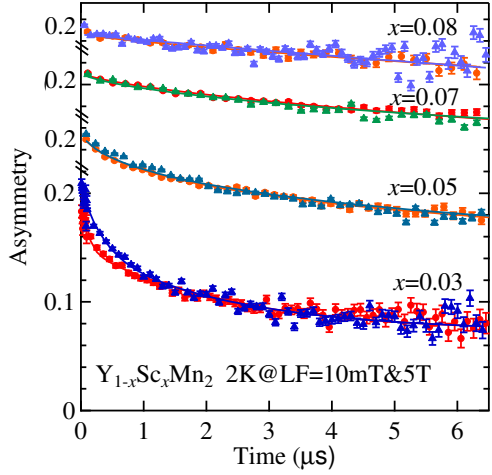}
\caption{(Color online)
Typical examples of \msr\ spectra [time-dependent asymmetry, $A(t)$] observed for \ymn\ at 2 K under two different longitudinal fields [$B_0=10$ mT (circles) and 5 T (triangles)], where spectra for $x\ge0.05$ are shifted vertically by 0.05 with each $x$ for clarity.  Solid curves show best fits using the function described in the text. 
Those for $x=0.03$ exhibit a slight dependence on $B_0$, from which the hyperfine parameter is estimated, whereas the spectra are less dependent on $B_0$ for $x\ge0.05$. 
}
\label{tspec}
\end{center}
\end{figure}

Figure \ref{tspec} shows some examples of time-dependent \msr\ spectra at a low temperature ($\simeq2$ K), where pairs of spectra obtained at two different longitudinal fields (LF, 10 mT and 5 T) are displayed for comparison\cite{Miyazaki:11}. The depolarization rate ($\lambda$) decreases with increasing Sc content $x$ and tends to approach an asymptotic value (as it exhibits little change between $x=0.07$ and 0.08).  
It is also noticeable that $\lambda$ is mostly independent of the magnitude of the longitudinal  
field ($B_0$) for $x\ge0.05$, while it shows a slight variation with $B_0$ for $x=0.03$.  These features can be readily understood from Eq.~(\ref{rfd}) introduced in the previous section.
We also note that Eq.~(\ref{rfd}) has been successfully applied to various types of magnetism including that of quasi-1D compounds \cite{Yamauchi:10}. Equation (\ref{rfd})  is modified to yield the fluctuation rate 
$$\nu\simeq \frac{(\delta_\mu^\parallel)^2k_BT\chi}{2N_A\mu_B^2\lambda}\pm\left[\left(\frac{(\delta_\mu^\parallel)^2k_BT\chi}{2N_A\mu_B^2\lambda}\right)^2-\omega_\mu^2\right]^{1/2}$$ 
from the experimental values of $\lambda$ and $\chi$, where the double sign corresponds to the two cases of $\nu>\omega_\mu$ ($+$) and $\nu<\omega_\mu$ ($-$) [see Fig.~\ref{ymn-cr}(c)]. The fact that the spectra in Fig.~\ref{tspec} are mostly independent of $\omega_\mu$ indicates that $\nu\gg\omega_\mu$, from which Eq.~(\ref{rfd}) is reduced to yield
\begin{equation}
\nu\simeq \frac{(\delta_\mu^\parallel)^2k_BT\chi}{N_A\mu_B^2\lambda}.\label{nulmd}
\end{equation}

\begin{figure}[t]
	\begin{center}
		\includegraphics[width=0.475\textwidth,clip]{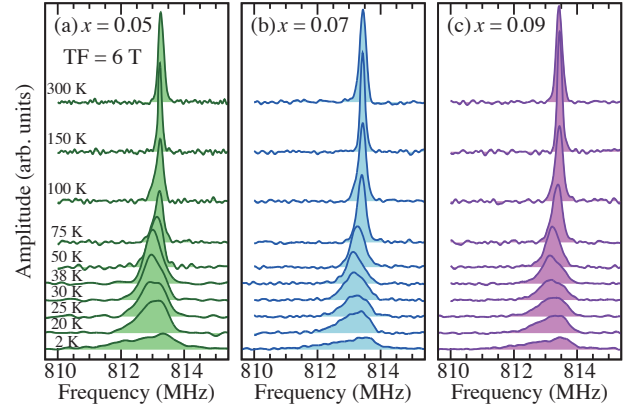}
		\caption{(Color online) FFTs of $\mu$SR spectra observed at various temperatures for Y$_{1-x}$Sc$_x$Mn$_2$ with (a) $x=0.05$, (b) 0.07, and (c) 0.09.}
	\label{ymn-fft}
	\end{center}
\end{figure}

The magnitude of $\delta_\mu^\parallel$ was determined as the gradient of the muon Knight shift ($K$) plotted against susceptibility (i.e., $dK/d\chi$ in the $K$-$\chi$ plot), for which additional \msr\ measurements under a high transverse field (HTF) were performed on  freshly synthesized samples with $x=0.05$, 0.07, and 0.09.\cite{Yamauchi:14} The fast Fourier transforms (FFTs) of the HTF-$\mu$SR spectra observed for these samples at various temperatures are shown in Fig.~\ref{ymn-fft}. These spectra can be used to monitor the density distribution of the local internal field at the muon site  [$P(\omega_\mu)$] via the relation 
\begin{equation}
\omega_\mu = \omega_0+ \gamma_{\mu}B_{\rm loc}, 
\end{equation}
where $B_{\rm loc}$ is the local field from nearby electrons.  The second moment of the local field distribution parallel to $B_0$ is directly related to the corresponding hyperfine parameter $\delta_\mu$ ($\equiv\delta_\mu^\perp$) as
\begin{equation}
\delta_\mu^2=\frac{1}{2}(\delta_\mu^\parallel)^2 \propto \gamma_{\mu}^2 \overline{(B_{\rm loc})_\parallel^2} 
\end{equation}
in the case of isotropic hyperfine fields, where $(B_{\rm loc})_\parallel$ is the component of $B_{\rm loc}$ parallel to $B_0$ and the overline denotes the mean value. The transverse relaxation rate is then given by
\begin{equation}
\lambda_\perp\simeq\left[\frac{2\delta_\mu^2\nu}{\nu^2+\omega_\mu^2}+\frac{\delta_\mu^2}{\nu}\right]\cdot \frac{k_BT\chi}{N_A\mu_B^2}.\label{lmdtf}
\end{equation}

\begin{figure}[tb]
\begin{center}
\includegraphics[width=0.5\textwidth]{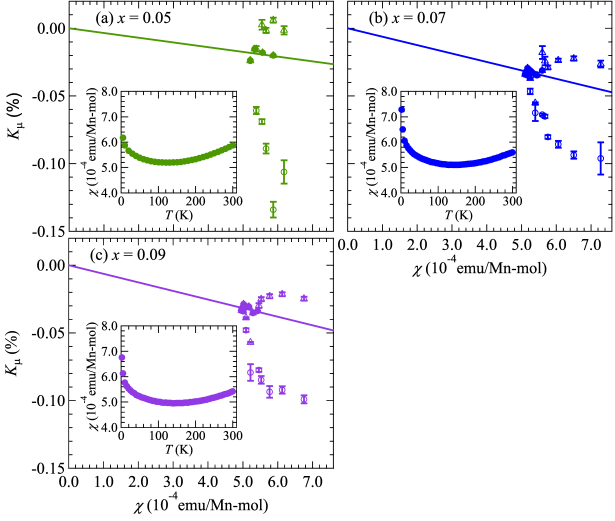}
\caption{(Color online)
$K$-$\chi$ plots obtained for \ymn\ with (a) $x=0.05$, (b) 0.07, and (c) 0.09, where the solid lines are the results of curve fitting for the data above $T^*\sim$100 K. Inset: temperature dependence of magnetic susceptibility ($\chi$) used in the $K$-$\chi$ plot.  
}
\label{Kchi}
\end{center}
\end{figure}

It is noticeable that the spectra become broad and asymmetric with decreasing $T$ below $T^*\sim100$ K, which seems in accordance with the onset of the Curie--Weiss behavior of $\chi$ (see Fig.~\ref{Kchi} inset).  The magnitude of the linewidth ($\lambda_\perp\simeq2$--5 MHz at 2 K) is much greater than that expected from Eq.~(\ref{lmdtf}) with $\delta_\mu$ determined by $dK/d\chi$. Such broadening was also reported in the cases of \lvo\cite{Koda:05} and \ymnz\ (see below). The nearly identical linewidth for different Sc contents $x$ shown in Fig. \ref{ymn-fft} indicates that the broadening at lower temperatures is not due to extrinsic effects such as crystallographic randomness owing to the Sc substitution or the formation of a spin-glass state observed for a smaller Sc content \cite{Mekata:00}. As is discussed in Sect.~\ref{anis}, we attribute this broadening to the strong spin fluctuation associated with geometrical frustration. 

Curve fitting of these spectra using the form
\begin{equation}
AG_x(t) = A_0\sum_{i=1}^m\exp(-\lambda_\perp^{(i)} t)\cos(\omega_\mu^{(i)} t+\phi_0) \label{TFana}
\end{equation}
yielded reasonable agreement with data for $T\ge T^*$ assuming one frequency component ($m=1$), where $A_0$ is the initial asymmetry, $\lambda_\perp$ is the transverse spin depolarization rate, and $\phi_0$ is the initial phase.  The Knight shift was then determined by
\begin{equation}
K_i=\frac{\omega_\mu^{(i)}-\omega_0}{\omega_0},\label{Kshift}
\end{equation}
where $\omega_0$ was determined by additional measurements on a reference sample of high purity silver.  Meanwhile, two frequency components ($m=2$) were incorporated to obtain satisfactory fits for the data below $T^*$.

Some examples of $K$-$\chi$ plots are shown in Fig.~\ref{Kchi}\cite{Yamauchi:14}. Considering the line broadening at lower temperatures, the hyperfine parameters are deduced from $dK/d\chi$ for $T\ge T^*$ to yield $\delta_\mu/2\pi=-26(3)$ MHz$/\mu_B$ ($x=0.05$), $-47(1)$ MHz$/\mu_B$ ($x=0.07$), and  $-49(1)$ MHz$/\mu_B$ ($x=0.09$), where the scattering among different $x$ is probably due to the residual influence of the line broadening (note that the relative variation of $K$ and $\chi$ is small for the relevant temperature region). However, their mean value [$\delta_\mu/2\pi=-41(2)$ MHz/$\mu_B$] is in perfect agreement with the calculated value assuming that muons are located at the 16$c$ site\cite{Mekata:00,Hartmann:90}, $|\delta_\mu/2\pi\mu_B|=$ 40.2 MHz/$\mu_B$. Although it is inferred from the $^1$H NMR
of YMn$_2$ that the hydrogen site is the 96$g$ site, which is slightly away from the 16$c$ site\cite{Fujiwara:87}, the calculated value of the muon hyperfine parameter ($\delta_\mu/2\pi= 78.6$ MHz/$\mu_B$) makes it unlikely that a muon occupies this site. 

The LF-\msr\ time spectra in Fig.~\ref{tspec} were analyzed by least-squares curve fitting to deduce $\lambda$  using 
\begin{eqnarray}
AG_z(t)&=&A_0G^{\rm KT}_{\rm z}(t)\cdot[(1-a_p)\exp\{-(\lambda t)^\beta\}+a_p]\\
&\simeq&[(1-a_p)\exp\{-(\lambda t)^\beta\}+a_p],\label{gzt}
\end{eqnarray}
where $G^{\rm KT}_{\rm z}(t)$ is the Kubo-Toyabe relaxation function, which is approximated by $G^{\rm KT}_{\rm z}(t)\simeq 1$ for high $B_0$, $\beta$ is the power, and $a_p$ is a constant term. Here, we introduce stretched exponential decay ($\beta\neq1$) to reproduce the deviation of the spin dynamics from that described by the model of spin correlation with a single value of $\delta_\mu^\parallel$ and/or $\nu$ at a given temperature \cite{Mekata:00,Johnston:05}. The presence of the term $a_p$ is clearly inferred in the case of $x=0.03$ from the leveling off of the time spectra for $t\ge4$ $\mu$s (see Fig.~\ref{tspec}; as is also needed for $x=0.05$). It is presumed that these deviations from single exponential decay are related to the strong line broadening observed under a transverse field. Apart from this ambiguity, excellent fits were obtained in all of the cases with $a_p$ fixed to the values deduced at 2 K. The solid curves in Fig.~\ref{tspec} represent the best fits obtained under these conditions\cite{Miyazaki:11}.

 \begin{figure}[tb]
\begin{center}
\includegraphics[width=0.4\textwidth]{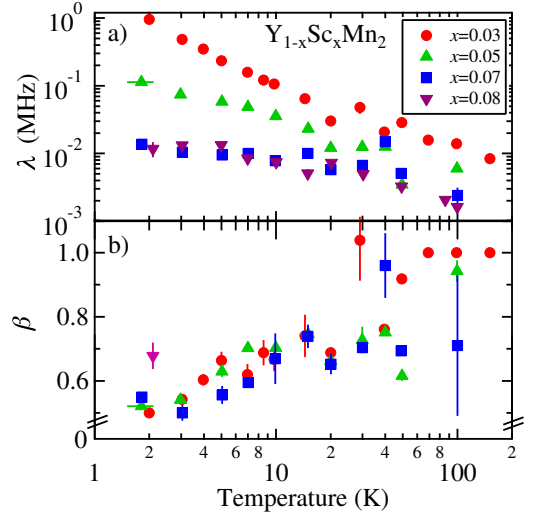}
\caption{(Color online)
Temperature dependences of (a) depolarization rate ($\lambda$) and power ($\beta$) for \ymn\ obtained by curve fitting of $A(t)$ for $x=0.03$ (circles), 0.05 (triangles), 0.07 (squares), and 0.08 (inverted triangles).  $\beta$ for $x=0.08$ is fixed at 2 K. 
}
\label{tparams}
\end{center}
\end{figure}

The temperature dependences of the parameters deduced by curve fitting are summarized in Fig.~\ref{tparams}.  Although $\beta$ varies only slightly with $x$, $\lambda$ exhibits a clear tendency of becoming less dependent on temperature, i.e.,
\begin{equation}
\lambda=\frac{1}{T_1}\propto T^0\:\:\:(T\le T^*), \label{ltpwr}
\end{equation}
 with increasing $x$. 
Considering the dependence of $\lambda$ on $\nu$ and $\chi$ in Eq.~(\ref{rfd}), this means that $\nu$ becomes linearly dependent on $T$ with increasing $x$.  
Meanwhile, the behavior of $\lambda$ for $x\rightarrow0.03$ is understood as the freezing of the Mn spin fluctuation because the transition to the quasistatic spin-glass state occurs in the sample with $x=0.03$ below $T_g\simeq 3$ K (where $T_g$ is defined as the peak muon depolarization rate under $B_0 = 10$ mT)\cite{Mekata:00}.  
The behavior of $\lambda$ observed for $x\ge0.07$ shows a distinct similarity to that for \lvo\ \cite{Koda:04}.

\begin{figure}[tb]
\begin{center}
\includegraphics[width=0.4\textwidth]{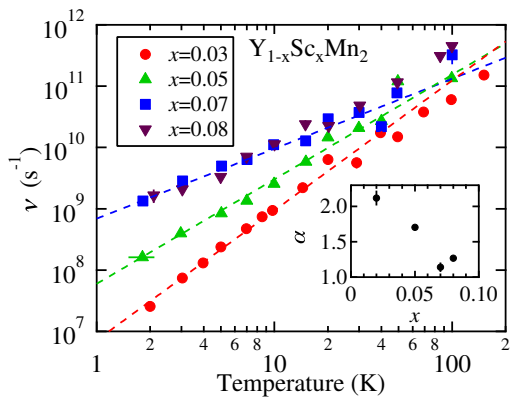}
\caption{(Color online)
Spin fluctuation rate ($\nu$) as a function of temperature and Sc content ($x$)  in \ymn.  Solid lines are the results of curve fitting using a power law ($\nu\propto T^\alpha$). Inset: $\alpha$ obtained by curve fitting vs $x$.
}
\label{ymn-nu}
\end{center}
\end{figure}

As shown in Fig.~\ref{ymn-nu}, the spin fluctuation rate in the samples with $x\ge0.07$ is in the range of $10^9$ --$10^{11}$ s$^{-1}$ for $T\le T^*$, while it shows a steeper reduction with decreasing temperature in those with $x\le0.05$. Although the use of stretched exponential decay in Eq.~(\ref{gzt}) prevents $\nu$ from being simply interpreted as a mean when $\beta<1$, $\nu$ serves as a ``characteristic frequency" that describes the spin dynamics on the basis of Eq.~(\ref{rfd}) \cite{Johnston:05}.
Solid lines are obtained by curve fitting using the power law 
\begin{equation}
\nu=c\cdot T^\alpha, \label{tpwr}
\end{equation}
with $c$ and $\alpha$ being free parameters.  As shown in the inset of Fig.~\ref{ymn-nu}, $\alpha$ exceeds 2 in the case of $x=0.03$, whereas it approaches unity ($\alpha\rightarrow1$) for $x\ge0.07$. 

Here, it is worth stressing that NMR and inelastic neutron scattering (INS) studies on the paramagnetic phase of \ymn\ conducted thus far have mostly been concerned with the relatively high energy part of the spin dynamics,  where they  demonstrated the presence of antiferromagnetic (AF) correlation with a characteristic frequency scale of $\nu_{\rm AF}\simeq10^{13}$--$10^{14}$ s$^{-1}$\cite{Nakamura:01,Zheng:99,Deportes:87,Shiga:88,Ballou:96}. In particular, a strong hyperfine field exerted on $^{55}$Mn nuclei  [corresponding to $\delta_\mu^\parallel/\sqrt{2}\simeq1.2$ GHz/$\mu_B$ in Eq.~(\ref{rfd})] drives the sensitive range of NMR up to such high frequencies \cite{Zheng:99}.
Interestingly, the latest INS study of a single-crystalline sample ($x=0.03$) revealed that the intensity centered at approximately ${\bm Q}_0=(1.25,1.25,0)$ (in reciprocal lattice units) exhibits anisotropic broadening along the Brillouin zone boundary, which is interpreted to be due to the degeneracy of states associated with geometrical frustration \cite{Ballou:96}.  Although this might be reminiscent of short-range correlations at a low energy, as similarly reported recently for \lvo\cite{Tomiyasu:14}, the details are yet to be clarified.

\subsection{YMn$_2$Zn$_{20-x}$In$_x$}

\begin{figure}
\begin{center}
\includegraphics[width=0.45\textwidth]{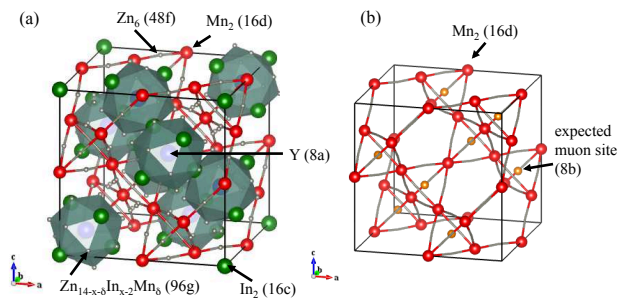}
\caption{(Color online) (a) Crystal structure of \ymzd, where Y, Mn, Zn, and In atoms are indicated by blue, red, white, and green spheres, respectively. The Zn at the 96$g$ site is partially substituted by In to stabilize the CeCr$_2$Al$_{20}$ structure. The actual compound exhibits a small amount ($\delta$) of excess Mn that occupies the 96$g$ site, where it substitutes for Zn. The Mn-Zn tetrahedra form a network similar to the pyrochlore lattice, where the local interaction between Mn atoms is relatively weaker than that in \ymn\ owing to the Zn atoms situated between the Mn atoms. (b) Expected muon site at the center of the Mn tetrahedra (8$b$ site) is shown.}\label{ymnz-cr}
\end{center}
\end{figure}

 One of the bottlenecks in the investigation of the $d$-electron HF state has been the limited number of candidate compounds that exhibit bulk electronic properties attributable to heavy-QP formation.  Recently, it has been reported that a ternary intermetallic compound, YMn$_2$Zn$_{20}$, exhibits a large Sommerfeld coefficient ($\gamma\ge 200$ mJ$\cdot$K$^{-2}$mol$^{-1}$) characteristic to the HF compounds\cite{Okamoto:10,Okamoto:12}. It crystallizes in the cubic CeCr$_2$Al$_{20}$ structure with the space group of $Fd\overline{3}m$ (see Fig.~\ref{ymnz-cr}), where the Y and Mn atoms respectively occupy the 8$a$ and 16$d$ sites, forming diamond and pyrochlore lattices that are common to the cubic Laves phase \ymn. Meanwhile, Zn atoms at the 16$c$, 48$f$, and 96$g$ sites are located between the other two atoms, so that the pyrochlore lattice composed of the Mn atoms is almost doubly expanded  in comparison with that in \ymn\ while keeping the tetrahedral symmetry. Although the compound with this ideal composition has not yet been obtained, the partial substitution of In for Zn is known to be effective for stabilizing the structure.

We have investigated the spin dynamics of Mn $d$-electrons by \msr\ under a zero/longitudinal field (ZF/LF) in a sample whose chemical composition is more precisely expressed as \ymzd\ with $x$=2.36\cite{Miyazaki:14}, where the influence of excess Mn appears to be minimal [$\delta$ = 0.11(1)]\cite{Okamoto:12}. The sample was a mosaic of single crystals glued with varnish on a sample holder made of high-purity silver (12 mm$\phi$ disc).
\msr\ measurements in the range of 300--4.2 K (using a $^4$He cryostat) were performed under an LF ($B_0 = 10$ mT) to quench the depolarization due to random local fields from nuclear magnetic moments. Those in the range of 50 K--0.3 mK (with a $^3$He cryostat) were performed under an LF ($B_0 = 395$ mT) to distinguish depolarization due to the pyrochlore (on-site) Mn from that due to the excess Mn, where the yield of these signals was estimated from the field dependence of the LF-\msr\ spectra at 0.3 K. 

\begin{figure}
\begin{center}
\includegraphics[width=0.475\textwidth]{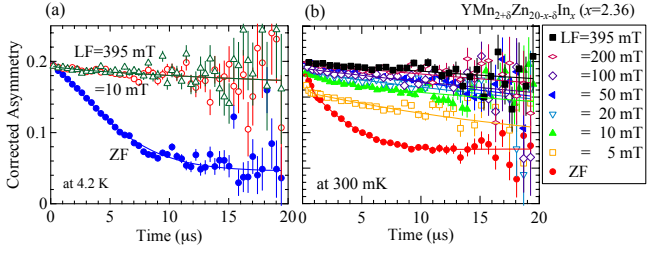}
\caption{(Color online) ZF/LF-\msr\ time spectra of \ymzd\ with $x$=2.36 ($\delta$=0.11) measured at (a) 4.2 K and (b) 300 mK. Solid curves are best fits using Eq.~(\ref{gzt}). The asymmetry of the spectrum under an LF of $B_0 = 395$ mT at 4.2 K is adjusted for comparison with the spectrum under an LF of $B_0 = 10$ mT. \label{ymnz-time}}
\end{center}
\end{figure}

Typical examples of ZF/LF-\msr\ spectra obtained at 4.2 and 0.3 K are shown in Fig.~\ref{ymnz-time}. The depolarization rate at 4.2 K is mostly independent of $B_0$  ($\lambda$ is unchanged between 10 and 395 mT), indicating that the spectra are in the limit of motional narrowing. These spectra were analyzed by curve fitting using Eq.~(\ref{gzt}), where $\beta=1$ and $a_p$ was replaced with $A_{\rm b}$ as the time-independent background mainly originating from muons stopped in the sample holder. 
The signal-to-background ratio ($A_{\rm 0}/A_{\rm b}$) was $\sim$3.8 for the $^4$He cryostat and $\sim$3 for the $^3$He cryostat over a time range of 0--20 $\mu$s, allowing the reliable deduction of $\lambda$ at small values ($\sim10^{-2} \mu$s$^{-1}$).

While we have successfully estimated the contribution of on-site Mn from the total magnetic susceptibility assuming two components of the Curie--Weiss term with one ($\chi_{\delta\rightarrow0}$) corresponding to the limit of $\delta=0$\cite{Okamoto:12}, it is often difficult to identify the contributions of the magnetic ions in question among those from impurities that also exhibit Curie--Weiss-like behavior.  The true contribution of the on-site Mn was confirmed by muon Knight shift measurements\cite{Miyazaki:15}, as was the case for \ymn.
The HTF-\msr\ spectra showed two frequency components with relative intensities of approximately 7 to 3, which was consistent with the values indicated from the partial asymmetries of the corresponding components in the LF-\msr\ spectra.  The time spectra were then analyzed by curve fitting using Eq.~(\ref{TFana}) assuming two components.

\begin{figure}[t]
	\begin{center}
		\includegraphics[width=0.4\textwidth]{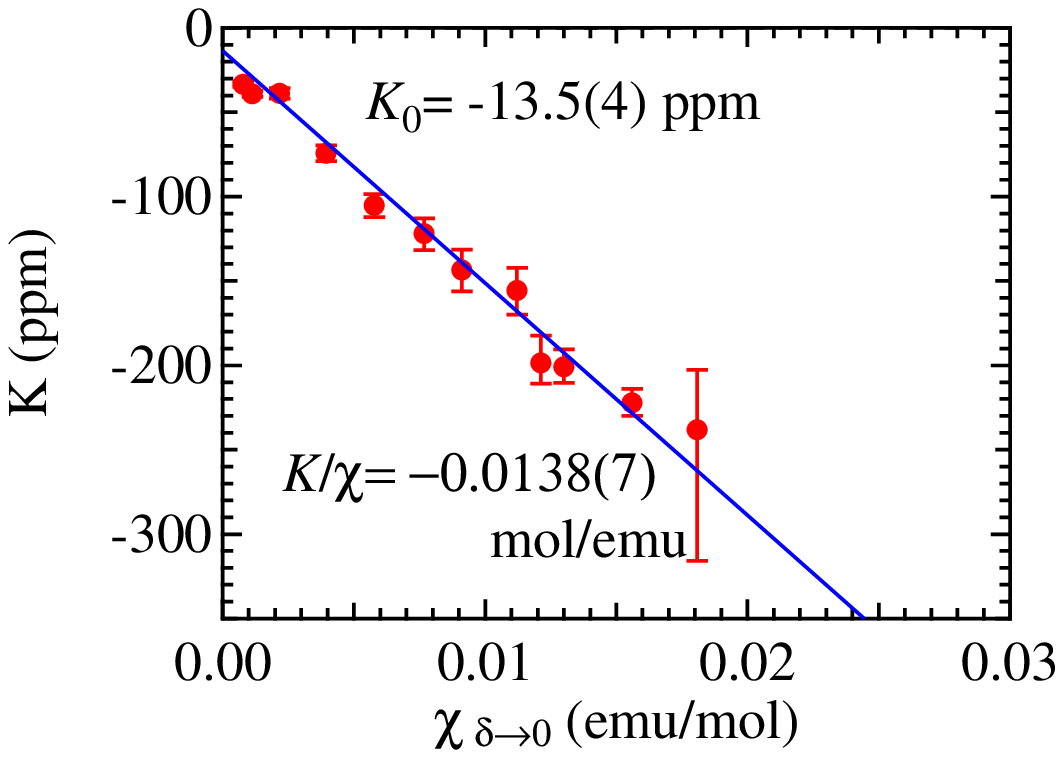}
		\caption{(Color online) Muon Knight shift ($K_1$) versus magnetic susceptibility ($\chi_{\delta\rightarrow0}$) for intrinsic pyrochlore Mn in \ymzd.}
	\label{ymnz-Kchi}
	\end{center}
\end{figure}

The muon Knight shift ($K_i$) for each signal was obtained by Eq.~(\ref{Kshift}) ($m=2$), where the ratio of the signal amplitudes was consistent with that observed in the ZF/LF-\msr\ spectra, in which the contributions of the pyrochlore (on-site) and excess Mn were readily identified.  This allowed us to unambiguously attribute the signal $K_1$ to the intrinsic pyrochlore Mn.  Curve fitting by the Curie--Weiss law yielded $|C|= 0.0036(3)$ emu$\cdot$K/mol and $\theta_{\rm W}=  -13(2)$ K, which showed nearly perfect agreement with $\chi_{\delta\rightarrow0}$ in the temperature dependence. Consequently, as shown in Fig.~\ref{ymnz-Kchi}, the $K_1$ versus $\chi_{\delta\rightarrow0}$ plot exhibited a linear relationship with $dK_1/d\chi_{\delta\rightarrow0}= -13.1(6)$ MHz/$\mu_B$ with a small offset of $K_0=13.5(4)$ ppm, indicating that $\chi_{\delta\rightarrow0}$ indeed originated from the on-site Mn.
Meanwhile, assuming that muons are located at the centers of Mn tetrahedra [8$b$ site, shown in Fig.~\ref{ymnz-cr}(b)],  $\delta_\mu^\parallel$ is calculated using the dipolar tensor for the Mn spins situated at the nearest- and next-nearest-neighboring tetrahedra to yield $|\delta_\mu^\parallel|=2\pi\times2.65$ MHz/$\mu_B$, which is in reasonable agreement with the experimental value.

\begin{figure}[t]
\begin{center}
\includegraphics[width=0.375\textwidth]{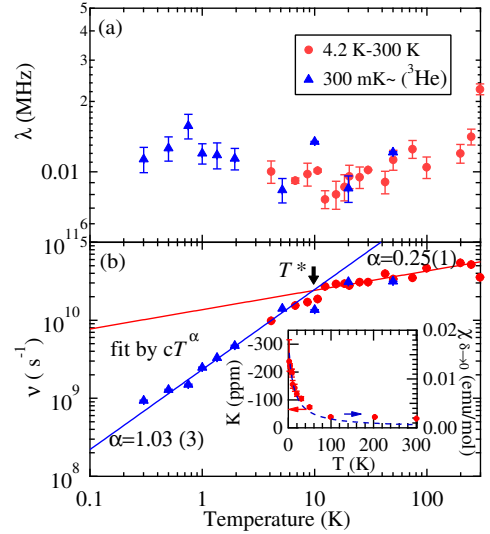}
\caption{(Color online) Temperature dependence of (a) longitudinal spin relaxation rate ($\lambda$) and (b) spin fluctuation rate ($\nu$) in \ymzd\ with $x$=2.36 ($\delta$=0.11). The solid curves in (b) are results of least-squares fitting using a power law ($\nu\propto T^\alpha$) for the range of temperatures above and below  $T^*$.}\label{ymnz-nu}
\end{center}
\end{figure}

The potential influence of excess Mn can be avoided by employing values of $\lambda$ obtained from data under an LF with magnitude greater than 100 mT.  As shown in Fig.~\ref{ymnz-nu}(a), $\lambda$ tends to level off ($\lambda\propto T^0$) for $T\le T^*$, which strongly resembles the case of \ymn\ with $x\ge0.07$.
Combining this result with $\chi_{\delta\rightarrow0}$ and $\delta_\mu$,  we obtain $\nu$ vs $T$ using Eq.~(\ref{rfd}) as shown in Fig.~\ref{ymnz-nu}(b).
The linear temperature dependence of $\nu$ emerges below $T^*\sim$10 K, confirming our previous result\cite{Miyazaki:14} with improved reliability.  Curve fitting using the power law [Eq.~(\ref{tpwr})] yields $\alpha=1.03(3)$\cite{Miyazaki:15}, which is again consistent with the case of \ymn. It is noticeable that $\nu$ exhibits a clear kink around $T^*\simeq10$ K, below which $\nu$ becomes linearly dependent on $T$. Thus, $T^*$ may be regarded as a crossover temperature below which the spin fluctuation is predominantly determined by the mechanism common to \ymn. 

We also note that $\lambda_\perp^{(1)}$  (the depolarization rate of the signal corresponding to the on-site Mn under a transverse field) was in the range of 1--2 MHz below $\sim$30 K whih was much greater than that expected from the magnitude of $\lambda$ [$\le0.02$ MHz, see Fig.~\ref{ymnz-nu}(a)], indicating the presence of additional line broadening due to the spin fluctuation associated with geometrical frustration in $\chi_{\rm local}$, as is also observed in \ymn\ and \lvo.

\subsection{LiV$_2$O$_4$}

Lithium vanadate is the only compound that exhibits HF behavior among the numerous metal oxides so far studied, and therefore it has attracted much attention since its discovery in the late '90s.\cite{Kondo:97,Urano:00}  The formation of a heavy-QP state below a characteristic temperature ($T_{\rm K}\simeq20$--30 K) is suggested by its large Sommerfeld coefficient ($\gamma\simeq 420$ mJ/mol$\cdot$K$^2$) and other bulk properties that are hallmarks of typical $f$-electron HF compounds.  Moreover, it has been inferred from the result of recent photoemission spectroscopy examination that a DOS peak slightly above $E_F$ develops for $T<T_{\rm K}$.\cite{Shimoyamada:06}

In our previous \msr\ study, we showed using a powder specimen of \lvo\ that the observed \msr\ spectra consisted of two components that could be distinguished by the response of the depolarization rate to an external magnetic field ($B_0$).\cite{Koda:04}   Furthermore, the component that exhibited the weaker dependence on $B_0$ (with longitudinal spin relaxation rate $\lambda_D$ and fractional yield $f\simeq0.4$) was mostly independent of temperature below $T^*\sim10^2$ K, from which we suggested that the corresponding fluctuation rate derived from the Redfield theory for the {\sl local} spin systems was also independent of temperature  ($\nu_D>10^9$ s$^{-1}$).  In contrast, the depolarization rate associated with the other signal ($\lambda_S$, with $1-f\simeq0.6$) was readily suppressed by $B_0$, which was ascribed to slowly fluctuating local magnetic moments ($\nu_S\sim10^6$--$10^7$ s$^{-1}$).  Although the occurrence of such phase separation was confirmed by a subsequent \msr\ study of high-quality single-crystalline samples,  the increased yield $f$ ($\simeq0.8$) implied that clarifying the origin of $\nu_D$ was essential to our understanding of the electronic state.\cite{Koda:05}

\begin{figure}[t]
\begin{center}
\includegraphics[width=0.375\textwidth]{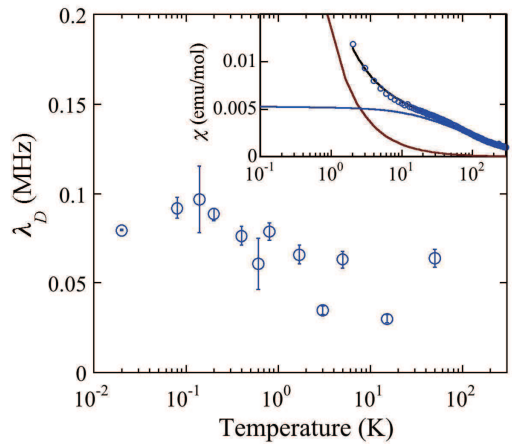}
\caption{(Color online)
Muon depolarization rate in \lvo\ reproduced using data in Ref.~\cite{Koda:04} for the component showing fast fluctuation ($\nu_D$).
Inset: Temperature ($T$) dependence of $\chi$ with $T$ having a logarithmic scale. Solid curves are results of curve fitting assuming two components, where the curve approaching 0.005 emu/mol is attributed to the intrinsic contribution of V ($\chi_{\rm V}$, see text for more details).
}
\label{lmd}
\end{center}
\end{figure}

 Figure \ref{lmd} shows the muon depolarization rate ($\lambda_D$) under a zero external field, deduced by curve fitting using the sum of two components with exponential damping:\cite{Koda:04} $$G_z(t)=f\exp(-\lambda_Dt)+(1-f)\exp(-\lambda_St).$$ 
 Although the data are scarce, particularly at higher temperatures, $\lambda_D$ is only weakly dependent on temperature with a tendency to level off with decreasing temperature, which is qualitatively similar to the behavior of $\chi_{\rm V}$ (Fig.~\ref{lmd}, inset). These features are remarkably similar to those observed for \ymn\ ($x\ge0.07$) and \ymnz.

A more appropriate evaluation of $\nu_D$ from $\lambda_D$ using Eq.~(\ref{rfd}) for an itinerant system was recently carried out.\cite{Kadono:12} Since the muon Knight shift data obtained from the previous \msr\ measurement were not sufficient to evaluate the temperature dependence of $K_i$ for the relevant component (particularly at low temperatures),\cite{Koda:04} we extracted it from the bulk susceptibility ($\chi$).
More specifically, considering the behavior of $\chi$ observed in single-crystalline samples, which tends to saturate at lower temperatures,\cite{Urano:00,Matsushita:05} we attributed the divergent behavior of $\chi$ at $T\rightarrow 0$ to unknown paramagnetic impurities (obeying the Curie law) and decomposed the data into two parts:
\begin{eqnarray}
\chi & = & \chi_{\rm V} + \chi_{\rm imp} \nonumber\\
          & = & \frac{C_{\rm V}}{T-\theta_{\rm W}} + \frac{C_{\rm imp}}{T},\label{chit}
\end{eqnarray}
where $\theta_{\rm W}$ is the Weiss temperature.  Curve fitting using Eq.~(\ref{chit}) yielded 
$C_{\rm V}=0.387(5)$ emu$\cdot$K/mol, $\theta_W=-74(2)$ K, and $C_{\rm imp}=0.0131(2)$ emu$\cdot$K/mol, implying that the behavior of $\chi_{\rm V}$ was in good agreement with that of single crystals [thus, we use $\chi=\chi_{\rm V}$ in Eq.~(\ref{rfd})].   Assuming that the Curie term originates from free V spins (V$^{3.5+}\sim$1.5$\mu_B$), the fractional yield of the impurity phase estimated as $C_{\rm imp}/T$ was 1.6\% of the total volume. This was much smaller than that of primary \msr\ signals (either $f$ or $1-f$ with $f\simeq0.4$), indicating that the paramagnetism of the impurity phase was irrelevant to the interpretation of \msr\ data. 

Another important quantity in Eq.~(\ref{rfd}) is $\delta_\mu^\parallel$. The muon Knight shift measurements of both powder and single-crystalline samples yielded $\delta_{\mu(D)}^\parallel\simeq0.5\pm0.2$ GHz/$\mu_B$, which corresponds to the component from which $\nu_D$ was deduced.\cite{Koda:04,Koda:05}  Apart from the large error due to the broad linewidth, this $\delta_{\mu(D)}^\parallel$ was in good agreement with the calculated $\delta_{\mu(D)}^\parallel$ of 0.143 GHz/$\mu_B$ for muons that occupy a site at the center of a cyclic vanadium hexamer (as inferred from the observed \msr\ linewidth due to {\sl nuclear} magnetic moments at high temperatures) and are subject to the magnetic dipolar fields from vanadium ions.\cite{Koda:04}  Since the experimental value of the hyperfine parameter had a large uncertainty due to the broad linewidth, we used the calculated value for the evaluation of $\nu_D$\cite{Kadono:12}.

\begin{figure}[t]
\begin{center}
\includegraphics[width=0.375\textwidth]{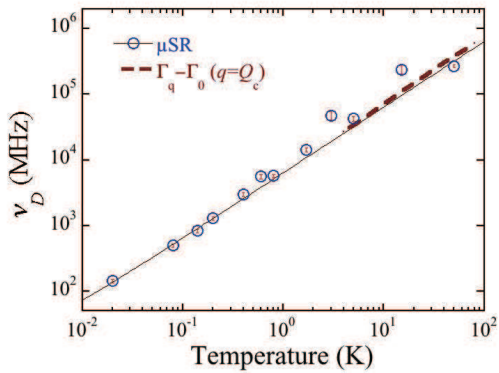}
\caption{(Color online)
Spin fluctuation rate ($\nu$) in \lvo\ as a function of temperature.  The thin solid line shows a linear $T$ dependence ($\nu\propto T$).   Inelastic neutron scattering data are also shown for comparison, where the dashed curve shows the linewidth $(\Gamma_q-\Gamma_0)/h$ at $Q_c=0.64$ \AA$^{-1}$ with $\Gamma_0\simeq1.5$ meV.\cite{Lee:01} }
\label{lvo-nu}
\end{center}
\end{figure}

The reevaluated $\nu_D$ is plotted in Fig.~\ref{lvo-nu} together with INS data,\cite{Lee:01}  where one can observe that $\nu_D$ lies on a straight line, indicating its proportional relationship with temperature ($\nu_D\propto T$) over a $T$ range of three decades below $T^*$.  This is again strikingly similar to the behavior of the spin fluctuation rate observed in \ymn\ and \ymnz\ at lower temperatures ($T<T^*$).  We also note that the corresponding low-energy excitation (also suggested in an earlier neutron scattering study\cite{Murani:04}) has been confirmed by a recent INS experiment on high-quality samples\cite{Tomiyasu:14}.

\section{Discussion}

\subsection{Spin fluctuation rate}
As mentioned earlier, the heavy QP mass is phenomenologically understood to originate from the sharp increase in the DOS slightly above the Fermi level and the associated flattening of the band dispersion, i.e., $D(E_F)\propto (m^*)^\sigma\rightarrow \infty$ (with $\sigma$ determined by the dispersion relation of electrons). In rare-earth compounds, such enhancement is induced by the conversion of local $f$-electron degrees of freedom into $D(E)$ by the Kondo effect, which is observed as a peak structure of $D(E)$ near $E_F$. The observation of such a structure  ($\sim$4 meV above $E_F$) in \lvo\ by photoemission spectroscopy\cite{Shimoyamada:06} appears to favor the Kondo mechanism established for $f$-electron compounds as a common microscopic origin of the HF behavior in the relevant compounds.  However, further attempts to obtain support for this scenario have been elusive.  A  theoretical model to project $d$-electron states (1.5 electrons per V$^{3.5+}$ ion) onto the Kondo model by splitting them into two sub-bands by electronic correlation had to introduce an unusually large Kondo coupling ($J_K\sim10^3$ K) to overcome the competing effect of Hund coupling.\cite{Anisimov:99,Kusunose:00} Our \msr\ study of a single-crystalline sample provided evidence against the formation of a spin-singlet state as it showed the presence of the local vanadium moments at low temperatures far below $T^*\simeq T_K$ (interpreted as the Kondo temperature), where the ``local" spins would disappear in this scenario.\cite{Koda:04,Koda:05}  From this view point, the importance of a highly symmetric crystal structure and the potential influence of geometrical frustration have been stressed by various authors, leading to a wide variety of theoretical models.\cite{Lacroix:01,Fulde:01,Shannon:02,Burdin:02,Hopkinson:02,Fujimoto:02,Tsunetsugu:02,Yamashita:03,Laad:03,Arita:07} 

The coexistence of slowly fluctuating local moments with the heavy-QP state  at low temperatures in \ymn\ and \ymnz\ supports the above-mentioned argument for \lvo.   The presence of local moments over the relatively long time range of $\nu^{-1}\simeq10^{-11}$--$10^{-9}$ s indicates that the conventional Kondo coupling (which virtually eliminates local spins over a time scale longer than $\nu_{\rm ex}^{-1} = h/J_{\rm cf} \sim 10^{-14}$--$10^{-13}$ s, where $J_{\rm cf}$ is the exchange energy between conduction electrons and $f$ electrons) is not in effect, thereby suggesting a different origin of the heavy-QP state in these compounds. 

According to a theoretical investigation of intersecting Hubbard chains as a model of the pyrochlore sublattice in \lvo, the low-energy part of the spin dynamics is predicted to be described by a spin-spin correlation whose relaxation rate is proportional to temperature\cite{Lee:03}. The behavior
\begin{equation}
 \nu\propto T,
\end{equation}
which is commonly observed in \lvo, \ymn\ ($x\ge0.07$, where the system is far from the spin-glass instability), and \ymnz\ at lower temperatures, is perfectly consistent with the above prediction, suggesting that the $t_{2g}$ orbitals associated with Mn/V atoms retain their 1D character at low temperatures (energies).
In particular, the spin fluctuation rate ($\nu_D$) in \lvo\ deduced from \msr\  is perfectly in line with the relaxation rate ($\Gamma_Q\propto T$ for $T<10^2$ K) observed over the low-energy region of the INS spectrum \cite{Lee:01}.  This implies that both \msr\ (sensitive over $0\le|{\bm q}|\le$ $\sim$1 \AA$^{-1}$) and INS probe common parts of the spin fluctuation spectrum of \lvo, suggesting that a similar situation is realized for \ymn\ and \ymnz.  Phenomenologically, this may be interpreted to mean that the heavy QPs develop in the manner $D(E_F)\propto (m^*)^{\sigma}\propto 1/T$ at lower temperatures.

Among the many theoretical models for the microscopic origin of the HF state in \lvo, that proposed by Fujimoto regards the quasi-1D character of the $t_{2g}$ bands associated with the pyrochlore lattice (consisting of intersecting chains of $t_{2g}$ orbitals) as an essential basis for the description of the electronic state as it is expected that the hybridization between the 1D bands will be strongly suppressed owing to the geometrical configuration (frustration)\cite{Fujimoto:02}. This model incorporates the hybridization as a perturbation to the 1D Hubbard bands, which yields an energy scale ($T^*$) that characterizes the crossover from 1D to 3D as the Fermi liquid state develops with decreasing temperature below $T^*$.  The calculated specific heat coefficient taking account of the latter as the leading correction to the self-energy yields a large value that is consistent with the experimentally observed values.  The progression of hybridization also induces the enhancement of the 3D-like spin correlation, which should appear as the enhancement of uniform susceptibility for $T<T^*$, while the spin fluctuation is dominated by the staggered component of 1D Hubbard chains.  

\subsection{Anomalous broadening of TF-\msr\ spectra}\label{anis}

\begin{figure}[b]
	\begin{center}
		\includegraphics[width=0.375\textwidth]{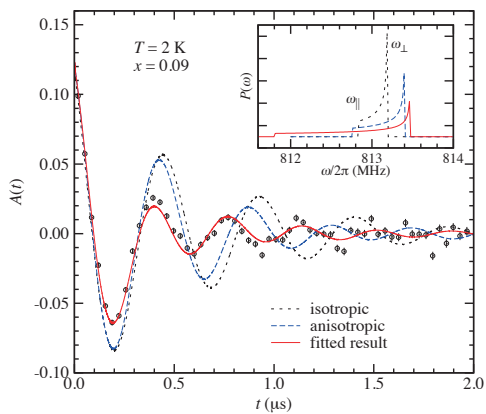}
		\caption{(Color online) Time dependent $\mu$SR spectra for \ysmn\ with $x$ = 0.09 obtained at 2 K under a transverse field of 6 T.  Dotted and broken curves are those calculated using $P(\omega)$ shown in the inset respectively assuming isotropic and anisotropic local spin susceptibility (see text for more details), and solid curves show the result of curve fitting with the uniaxial $P(\omega)$ plus additional line broadening. The data points and curves are displayed on a rotating reference frame with a frequency of 811 MHz. }
	\label{ymn-pp}
	\end{center}
\end{figure}

Here we discuss the origin of the strong broadening of the TF-\msr\ linewidth observed in \ymn\ at lower tempertures ($T<T^*$), which is also expected to be relevant to the other two compounds.  
Since the muon site has a $\bar{3}m$ point symmetry for the cubic $Fd\bar{3}m$ structure, 
the presumed muon site suggests a powder pattern for $P(\omega)$ that is characterized by two singularities (edges) at $\omega_\perp$ and $\omega_\parallel$ due to the uniaxially anisotropic dipolar fields\cite{Slichter},  
\begin{equation}
P(\omega) = \frac{1}{2 \sqrt{(\omega - \omega_\perp)(\omega_\parallel - \omega_\perp)}}
\label{eq:P}
\end{equation}
(see the inset of Fig.~\ref{ymn-pp}).
Accordingly, we attempt to reproduce TF-$\mu$SR time spectra by curve fittting using the following form:  
\begin{equation}
G_x(t) =  \exp[-(\sigma t)^{\beta}] \int_{\omega_\parallel}^{\omega_\perp} P(\omega_\mu) \cos(\omega_\mu t + \phi_0) d \omega_\mu,
\label{eq:uniax}
\end{equation}
where stretched exponential damping was incorporated to consider the additional line broadening due to spin fluctuation.

Figure \ref{ymn-pp} shows result of such curve fitting for a time spectrum obtained at 2 K in the time domain with the inset  
showing the calculated $P(\omega)$ for (i) isotropic local spin susceptibility (black dotted curve) and (ii) assuming in-plane anisotropy (blue dashed curve), where the Mn moments are allowed to fluctuate within the easy plane perpendicular to the threefold rotation-inversion axis on the Mn site (which is parallel to the [111] direction). The latter means that the local spin susceptibility probed by a muon strongly depends on the field direction. Then, the local spin susceptibility becomes considerably greater than the isotropic case when the external field is parallel to the principal axis, and thereby $\chi$ in Eq.~(\ref{lmdtf}) should be replaced with the anisotropic $\chi_{\rm local}$ (thus enhancing $\lambda_\perp$).   The main panel of Fig.~\ref{ymn-pp} shows the corresponding $A(t)=A_0G_x(t)$ (black dotted and blue broken curves) calculated for the respective $P(\omega)$ using the parameters $A_0$, $\lambda$, $\beta$, and $\phi_0$ obtained from the curve fitting using Eq.~(\ref{eq:uniax}). One can observe that the spectrum is not reproduced by the two types of calculated curve and that further broadening is required. This clearly indicates the occurrence of excess line broadening, which is not explained by the anisotropy of the Knight shift. We note that a similar result was obtained for the TF-\msr\ time spectra in \lvo.  Such broadening may be a manifestation of strong spin fluctuation due to the geometrical frustration. 

In any case, the increase in $\lambda_\perp$ for $T<T^*$ is commonly observed in \ymn, \ymnz, and \lvo, and it is expected to be an important clue to understanding the HF-like behavior of these pyrochlore antiferromagnets.

\section{Summary and Conclusion}
We have shown that the longitudinal spin relaxation rate in the $d$-electron HF-like compounds commonly exhibits a  tendency to level off, $\lambda=1/T_1\propto T^0$, below a characteristic temperature $T^*$.  This implies that the spin fluctuation rate becomes linearly dependent on temperature, $\nu\propto T$, in these compounds for $T<T^*$. In particular, such behavior in \lvo\ is consistent with the implications of INS experiments, providing a basis for the coherent understanding of the low-energy spin dynamics through the theoretical model of intersecting 1D Hubbard chains that simulates a pyrochlore lattice.  The persistent quasi-1D spin dynamics coexists with the enhanced local susceptibility at lower temperatures ($T<T^*$), which is also common to two other $d$-electron HF systems, \ymn\ and \ymnz.  These observations strongly indicate that a geometrically constrained $t_{2g}$ band is the primary requirement for the formation of heavy QPs, with  the 1D-to-3D crossover as a possible mechanism of effective mass enhancement.

\vspace{1ex}
\begin{acknowledgments}
The results quoted in this review were achieved under collaborations with T. Yamazaki, Y. Tabata, and H. Nakamura for \ymn, with Y. Okamoto and Z. Hiroi for \ymnz, and with H. Ueda, C. Urano, S. Kondo, M. Nohara, H. Takagi, Y. Matsushita, and Y. Ueda for \lvo. We would like to thank M. Hiraishi, T. Masuda, A. Koda, K. M. Kojima, K. Ohishi, I. Kawasaki, I. Watanabe, and W. Higemoto for their active involvement in \msr\ experiments. We also thank K. Tomiyasu, S. Fujimoto, H. Tsunetugu, and Y. Kuramoto for helpful discussions.  The work was partly supported by the Inter-University Research Program of Institute of Materials Structure Science, KEK (Proposal Nos.~2011A0011, 2012A0051, 2013A0089, 2014A0238).

\end{acknowledgments}


\begin{thebibliography}{9}
\bibitem{Wada:89} H. Wada, M. Shiga, and Y. Nakamura, Physica B {\bf 161}, 197 (1989).

\bibitem{Fisher:93} R. A. Fisher, R. Ballou, J. P. Emerson, E. Lelievre-Berna, and N. E. Philips, 
Int. J. Mod. Phys. B {\bf 7}, 830 (1993). 

\bibitem{Kondo:97} S. Kondo, D. C. Johnston, C. A. Swenson, F. Borsa, A. V. Mahajan, L. L. Miller, T. Gu, A. I. Goldman, M. B. Maple, D. A. Gajewski, E. J. Freeman, N. R. Dilley, R. P. Dickey, J. Merrin, K. Kojima, G. M. Luke, Y. J. Uemura, O. Chmaissem, and J. D. Jorgensen, Phys. Rev. Lett. {\bf 78}, 3729 (1997).

\bibitem{Urano:00} C. Urano, M. Nohara, S. Kondo, F. Sakai, H. Takagi, T. Shiraki, and T. Okubo, 
Phys. Rev. Lett. {\bf 85},1052 (2000).

\bibitem{Miyazaki:11} M. Miyazaki, R. Kadono, M. Hiraishi, T. Masuda, A. Koda, K. M. Kojima, T. Yamazaki, Y. Tabata, and H. Nakamura, J. Phys. Soc. Jpn. {\bf 80}, 063707 (2011).

\bibitem{Miyazaki:14} M. Miyazaki, R. Kadono, M. Hiraishi, I. Yamauchi, A. Koda, K. M. Kojima, I. Kawasaki, I. Watanabe, Y. Okamoto, and Z. Hiroi, J. Phys.: Conf. Ser. {\bf 551}, 012019 (2014).

\bibitem{Koda:04} A. Koda, R. Kadono, W. Higemoto, K. Ohishi, H. Ueda, C. Urano, S. Kondo, M. Nohara, and H. Takagi, 
Phys. Rev. B {\bf 69}, 012402 (2004).

\bibitem{Kadono:12} R. Kadono, A. Koda, W. Higemoto, K. Ohishi, H. Ueda, C. Urano, S. Kondo, M. Nohara, and H. Takagi, J. Phys. Soc. Jpn. {\bf 81}, 014709 (2012).

\bibitem{Koda:05} A. Koda, R. Kadono, K. Ohishi, S. R. Saha, W. Higemoto, Y. Matsushita, and Y. Ueda, J. Phys.: Condens. Matter {\bf 17}, L257 (2005).

\bibitem{Yamauchi:14} I. Yamauchi, M. Miyazaki, M. Hiraishi, A. Koda, K. M. Kojima, R. Kadono, and H. Nakamura,  unpublished.

\bibitem{Schenck} 
A.~Schenck, {\it Muon Spin Rotation Spectroscopy: Principles and Applications in Solid State Physics} (Adam Hilger, Bristol, 1985) Vol. 125.

\bibitem{Moriya:56} T. Moriya, 
Prog. Theor. Phys. {\bf 16}, 23 (1956); Prog. Theor. Phys. {\bf 16},  641 (1956).

\bibitem{Ballou:87} R. Ballou, J. Deportes, R. Lemaire, Y. Nakamura, and B. Ouladdiaf, 
J. Magn.  Magn. Mater. {\bf 70}, 129 (1987).

\bibitem{Nakamura:88} H. Nakamura, H. Wada, K. Yoshimura, M. Shiga, Y. Nakamura, J. Sakurai, and Y. Komura, 
J. Phys. F: Met. Phys. {\bf 18}, 981 (1988).

\bibitem{Yamauchi:10} See, for example, I. Yamauchi, M. Itoh, T. Yamauchi, J. Yamaura, and Y. Ueda, J. Phys.: Conf. Ser. {\bf 200}, 012234 (2010).

\bibitem{Mekata:00}  M. Mekata, T. Asano, H. Nakamura, M. Shiga, K. M. Kojima, G. M. Luke, A. Keren, W. D. Wu, M. Larkin, Y. J. Uemura, S. Dunsinger, and M. Gingras, 
Phys. Rev. B {\bf 61}, 4088 (2000).

\bibitem{Hartmann:90}  O. Hartmann, R. W\"appling, K. Aggarval, L. Asch, A. Kratzer, G.~M. Kalvius, F.~J.~Litterst, A. Yaouanc, P. Dalmas de R\'eotier, B. Barbara, F.~N.~Gygax, B. Hitti, E. Lippelt, and A. Schenck, 
Hyperfine Interact. {\bf 64}, 711 (1990).

\bibitem{Fujiwara:87} K. Fujiwara, K. Ichinose, H. Nagai, and A. Tsujimura, 
J. Magn. Magn. Mater. {\bf 70}, 184 (1987).

\bibitem{Johnston:05} D.~C. Johnston, S.~H. Baek, X. Zong, F. Borsa, J. Schmalian, and S. Kondo, Phys. Rev. Lett. {\bf 95}, 176408 (2005).

\bibitem{Shiga:88} M. Shiga, H. Wada, Y. Nakamura, J. Deportes, B. Ouladdiaf, and K.~R.~A. Ziebeck, 
J. Phys. Soc. Jpn. {\bf 57},  3141 (1988).

\bibitem{Nakamura:01} H. Nakamura  and M. Shiga, 
J. Alloys  Compd.  {\bf 326}, 157 (2001).

\bibitem{Zheng:99} G.-q. Zheng, K. Nishikido, K. Ohonishi, Y. Kitaoka, K. Asayama, and R. Hauser, 
Phys. Rev. B {\bf 59}, 13973 (1999).

\bibitem{Deportes:87} J. Deportes, B. Ouladdiaf, and K.~R.~A. Ziebeck, 
J. Magn. Magn. Mater. {\bf 70}, 14  (1987).

\bibitem{Ballou:96} R. Ballou, E. Leli\'evre-Berna, and B. F\aa k, Phys. Rev. Lett. {\bf 76}, 2125 (1996).

\bibitem{Tomiyasu:14} K. Tomiyasu, K. Iwasa, H. Ueda, S. Niitaka, H. Takagi, S. Ohira-Kawamura, T. Kikuchi,
Y. Inamura, K. Nakajima, and K. Yamada, Phys. Rev. Lett. {\bf113},  236402 (2014).

\bibitem{Okamoto:10} Y. Okamoto, T. Shimizu, J. Yamaura, Y. Kiuchi, and Z. Hiroi, 
J. Phys. Soc. Jpn. {\bf 79}, 093712 (2010).

\bibitem{Okamoto:12} Y. Okamoto, T. Shimizu, J. Yamaura, Y. Kiuchi, and Z. Hiroi, J. Solid State Chem. {\bf 191},  246 (2012).

\bibitem{Miyazaki:15} M. Miyazaki, R. Kadono, M. Hiraishi, I. Yamauchi, A. Koda, K. M. Kojima, I. Kawasaki, I. Watanabe, Y. Okamoto, and Z. Hiroi, unpublished.

\bibitem{Shimoyamada:06} A. Shimoyamada, S. Tsuda, K. Ishizaka, T. Kiss, T. Shimojima, T. Togashi, S. Watanabe, C. Q. Zhang, C. T. Chen, Y. Matsushita, H. Ueda, Y. Ueda, and S. Shin, 
Phys. Rev. Lett. {\bf 96}, 026403 (2006).

\bibitem{Matsushita:05} Y. Matsushita, H. Ueda, and Y. Ueda, Nat. Mater. {\bf 4}, 845 (2005).

\bibitem{Lee:01} S.-H. Lee, Y. Qiu, C. Broholm, Y. Ueda, and J.~J.~Rush, 
Phys. Rev. Lett. {\bf 86}, 5554  (2001).

\bibitem{Murani:04} A. P. Murani, A. Krimmel, J. R. Stewart, M. Smith, P. Strobel, A. Loidl, and A. Ibarra-Palos, J. Phys.: Condens. Matter {\bf 16}, S607  (2004).

\bibitem{Anisimov:99} V. I. Anisimov, M. A. Korotin, M. Z\"olfl, T. Pruschke, K. Le Hur, and T. M. Rice, Phys. Rev. Lett. {\bf 83},  364 (1999).

\bibitem{Kusunose:00} H. Kusunose, S. Yotsuhashi, and K. Miyake, Phys. Rev. B {\bf 62}, 4403 (2000).

\bibitem{Lacroix:01} C. Lacroix, Can. J. Phys. {\bf 79}, 1469 (2001).

\bibitem{Fulde:01} P. Fulde, A. N. Yaresko, A. A. Zvyagin, and Y. Grin, Europhys. Lett. {\bf 54}, 779 (2001).

\bibitem{Shannon:02} N. Shannon, Eur. Phys. J. B {\bf 27}, 527 (2002).

\bibitem{Burdin:02} S. Burdin, D. R. Grempel, and A. Georges, Phys. Rev. B
{\bf 66}, 045111 (2002).

\bibitem{Fujimoto:02} S. Fujimoto, Phys. Rev. B {\bf 65}, 155108 (2002).

\bibitem{Hopkinson:02} J. Hopkinson and P. Coleman, Phys. Rev. Lett. {\bf 89}, 267201 (2002).

\bibitem{Tsunetsugu:02} H. Tsunetsugu, J. Phys. Soc. Jpn. {\bf 71}, 1844 (2002).

\bibitem{Yamashita:03} Y. Yamashita and K. Ueda, Phys. Rev. B {\bf 67}, 195107 (2003).

\bibitem{Laad:03} M. S. Laad, L. Craco, and E. M\"uller-Hartmann, Phys. Rev.
B {\bf 67}, 033105 (2003).

\bibitem{Arita:07} R. Arita, K. Held, A.V. Lukoyanov, and V. I. Anisimov,
Phys. Rev. Lett. {\bf 98}, 166402 (2007).

\bibitem{Lee:03} J.~D. Lee, Phys. Rev. B {\bf 67}, 153108 (2003).

\bibitem{Slichter} See, for example, C. P. Slichter, {\it Principles of Magnetic Resonance}, 3rd Edition, (Springer-Verlag, New York, 1990) 3rd ed.

\end{thebibliography}
\end{document}